\documentclass{webofc}
\usepackage[varg]{txfonts}
\usepackage{hyperref}
\usepackage{url}
\usepackage{amsmath,amssymb,mathtools}
\usepackage{slashed}
\usepackage{bm}
\usepackage{orcidlink}
\usepackage{bbm}
\usepackage{float}
\usepackage{graphicx}
\usepackage{microtype}

\hypersetup{colorlinks=true,citecolor=blue,urlcolor=blue,linkcolor=blue}
\begin{document}
\title{Removing cutoff artifacts in the NJL model by a Renormalization Group consistent treatment}

\author{\firstname{Hosein} \lastname{Gholami}\orcidlink{0009-0003-3194-926X}\inst{1}\fnsep\thanks{\email{mohammadhossein.gholami@tu-darmstadt.de }} \and
        \firstname{Marco} \lastname{Hofmann}\orcidlink{0000-0002-4947-1693}\inst{1}
        \fnsep
        \thanks{\email{marco.hofmann@tu-darmstadt.de}}
        \and
        \firstname{Michael} \lastname{Buballa}\orcidlink{0000-0003-3747-6865}\inst{1,2}\fnsep\thanks{\email{michael.buballa@tu-darmstadt.de}}
}

\institute{Technische Universit{\"a}t Darmstadt, Fachbereich Physik, Institut f{\"u}r Kernphysik,
Theoriezentrum, Schlossgartenstr. 2, D-64289 Darmstadt, Germany
\and
Helmholtz Forschungsakademie Hessen f{\"u}r FAIR (HFHF),
GSI Helmholtzzentrum f{\"u}r Schwerionenforschung,
Campus Darmstadt, D-64289 Darmstadt, Germany
          }

\abstract{We summarize how a renormalization-group (RG)-consistent treatment removes well-known artifacts in NJL-model descriptions of color-superconducting quark matter. We introduce two RG-consistent schemes, "minimal" and "massless", and present analytic solutions for the diquark gap at \(T=0\) and for the phase boundary \(T_c(\mu)\) in symmetric massless matter, representing the high-density limit of the model. We compare the pairing gaps, 
phase diagram, and speed of sound with results obtained using conventional regularization. 
}
\maketitle
\section*{Introduction}
\label{intro}
Nambu--Jona-Lasinio (NJL)-type models are widely used to study strong-interaction matter at moderately high density regions, which are phenomenologically interesting for compact-star physics but not accessible by lattice QCD calculations or perturbation theory. They are particularly well suited to investigate chiral symmetry breaking and color superconductivity (CSC) in a single, consistent framework. As shown in Ref.~\cite{Ruester:2005jc}, the phase structure of the NJL model in three-flavor, locally neutral quark matter exhibits a rich phase structure including several CSC phases. At intermediate densities, matter is in a two-flavor color superconducting (2SC) phase in which up and down quarks of two colors (red and green) are paired, whereas at high densities the color-flavor locked (CFL) phase, in which quarks of all flavors and colors are paired, is preferred. In addition, there can appear other color-superconducting phases such as the dSC phase ($ud$ and $ds$ pairing) and the uSC phase ($ud$ and $us$ pairing).

\begin{figure}[H]
\centering
\includegraphics[width=6.3cm,clip]{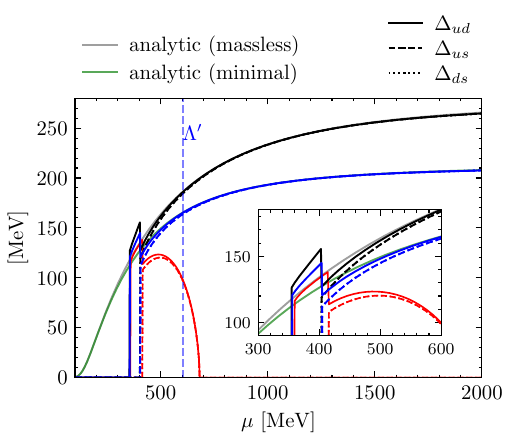}
\includegraphics[width=6.3cm,clip]{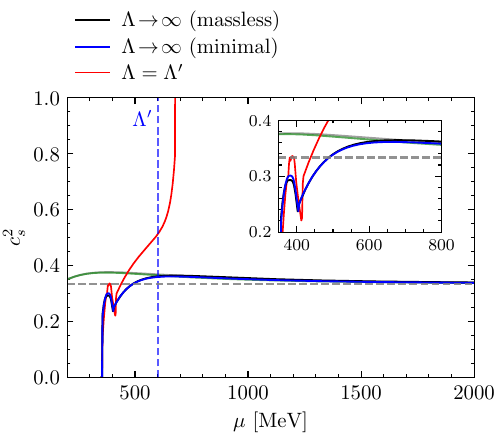}
\caption{Left: $T=0$ diquark gaps $\Delta_{ud}$ and $\Delta_{us}=\Delta_{ds}$ versus quark chemical potential $\mu$ in locally neutral matter. We compare the  conventional regularization (red) and the RG-consistent minimal (blue) and massless (black) scheme  with the analytic results
for massless, symmetric CFL matter in the minimal (green) and massless (gray) scheme, see Eq.~\eqref{eq:gapeqminimal} and the following discussion.
Right: comparison of the speed of sound squared $c_s^2$ (in units of the speed of light) with Eq.~\eqref{eq:cs2}.
The analytic curves  for massless and minimal scheme are very similar and overlap in the right plot.
}
\label{fig:gaps}
\end{figure}

Despite all this, the NJL model is a non-renormalizable theory. In the conventional approach, integrals are regularized with a single 3d momentum cutoff $\Lambda' \simeq 600\,$MeV, which is fitted to vacuum observables. This leads to cutoff artifacts as soon as the quark chemical potential $\mu$ or the temperature $T$ become comparable in size to the cutoff. These well-known artifacts include:
     (i) The diquark gaps decrease with increasing quark chemical potential and eventually vanish (see red curves in Fig.~\ref{fig:gaps} left).
(ii) The speed of sound becomes acausal and diverges (see red curves in Fig.~\ref{fig:gaps} right).
(iii) The critical temperature $T_c$ decreases with increasing chemical potential (red curves in Fig.~\ref{fig:pd}).
(iv) The melting pattern of the CFL phase disagrees with Ginzburg--Landau theory \cite{Iida:2003cc} (red curves in Fig.~\ref{fig:pd}).

In this work we implement a renormalization group (RG) consistent treatment for the three-flavor NJL model with color superconductivity in order to mitigate these cutoff artifacts.
\section*{Renormalization group consistent treatment}
RG consistency states that the full quantum effective action $\Gamma$ must not depend on the UV scale $\Lambda$ at which the classical action is initialized: $\Lambda \frac{d \Gamma}{d \Lambda}=0$. Based on this concept, the authors of Ref.~\cite{Braun:2018svj} suggested an RG-consistent treatment for mean-field models.
 In our NJL model, we  achieve such an RG-consistent action by integrating the vacuum flow up to a high scale $\Lambda \gg \mu,T,\Lambda'$ before integrating the medium flow to the infrared, for details see Ref.~\cite{Gholami:2024diy}. In the presence of nonzero diquark gaps, there are medium divergences 
$\sim\sum_{\alpha a,\beta b}\mu^2_{\alpha a,\beta b}\,|\Delta_{\alpha a,\beta b}|^2\ln\Lambda$, 
with flavor indices $\alpha,\beta\in\{u,d,s\}$, color indices $a,b\in\{r,g,b\}$, 
and $\mu_{\alpha a,\beta b}$ is the average chemical potential of the pair $\Delta_{\alpha a,\beta b}$.
To cancel these divergences, we subtract appropriate counterterms. In this context, we define the \emph{massless} and the \emph{minimal} scheme by demanding different correlators of the effective action vanish at RG scale $\Lambda'$(see Refs.~\cite{Gholami:2024diy,Gholami:2025afm} for details).
As pointed out in Ref.~\cite{Gholami:2024diy} and demonstrated in Ref.~\cite{Gholami:2025afm} within an RG-consistent two-flavor Quark--Meson--Diquark model, the minimal scheme can be interpreted as a wave-function renormalization of the diquark fields. Moreover, it was shown that the minimal scheme preserves the BCS relation $T_c^{2SC}\simeq 0.567 \Delta_{ud}(T=0)$ for symmetric matter, whereas the massless scheme does not. In the following section, we compare the conventional (3d-cutoff) regularization with the RG-consistent minimal and massless schemes in the NJL model.
Our comparison focuses on the zero-temperature diquark condensate, the speed of sound, and the $T$–$\mu$ phase diagram.
\section*{Numerical results}
We solve the gap equations and neutrality conditions of the model numerically using the same parameters as in Refs.~\cite{Ruester:2005jc,Gholami:2024diy}. The original NJL cutoff is $\Lambda'=602.3~\mathrm{MeV}$. For the RG-consistent setup, we take the RG scale $\Lambda$ numerically large (effectively $\Lambda\!\to\!\infty$). 

The left panel of Fig.~\ref{fig:gaps} shows the $T=0$ diquark gaps as functions of the quark chemical potential $\mu$. We compare three treatments: (i) a \emph{conventional} regularization with the higher RG scale fixed to the model cutoff, $\Lambda=\Lambda'$, (ii) the RG-consistent \emph{minimal} scheme, and (iii) the RG-consistent \emph{massless} scheme. A first-order transition into the 2SC phase occurs around $\mu\simeq 350~\mathrm{MeV}$ (onset of $\Delta_{ud}\neq0$), followed at higher $\mu$ by a first-order transition into the CFL phase with all gaps nonzero. In the conventionally regularized model, the diquark gaps drop to zero shortly after the cutoff is reached. By contrast, for the RG-consistent minimal and massless schemes the gaps approach the asymptotic values 
$\bar\Delta^{\text{(min)}}(\mu\!\to\!\infty)=213.3~\mathrm{MeV}$
and 
$\bar\Delta^{\text{(massless)}}(\mu\!\to\!\infty)=279.3~\mathrm{MeV}$, respectively (see analytic results in the next section).

The right panel of Fig.~\ref{fig:gaps} shows the speed of sound $c_s^2$ for the three schemes. In the conventional regularization $c_s^2$ diverges once the diquark gaps vanish, consistent with the behavior discussed in Ref.~\cite{Gholami:2025afm}. In the RG-consistent treatments, $c_s^2$ approaches the conformal limit $1/3$ from above. A first maximum appears in the 2SC phase; for the chosen parameters it is sub-conformal and decreases towards the CFL transition due to the onset of strange quarks in the system.

\begin{figure}[H]
\sidecaption
\centering
\includegraphics[width=9.62cm,clip]{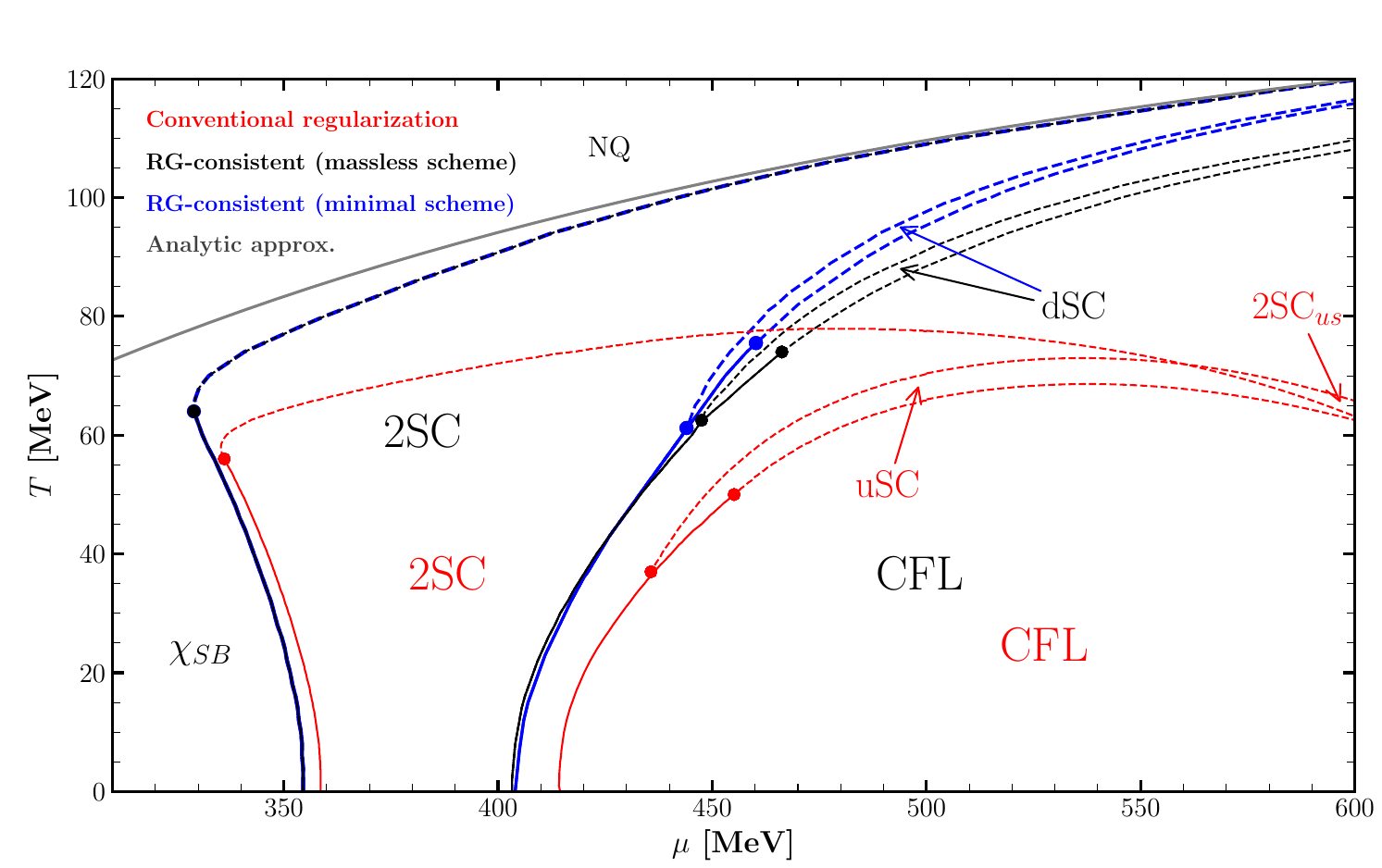}
\caption{Phase diagram of neutral color-superconducting matter for the different schemes presented in this work. Solid lines indicate first-order phase transitions and dashed lines indicate second-order transitions. The analytic curve is the RG-consistent symmetric CFL critical temperature in Eq.~\eqref{eq:tcNJL2}.}
\label{fig:pd}
\end{figure}

Fig.~\ref{fig:pd} displays the phase diagram for all schemes. In agreement with Ginzburg--Landau theory \cite{Iida:2003cc}, the RG-consistent model realizes the expected high-density melting sequence $\mathrm{CFL}\to\mathrm{dSC}\to\mathrm{2SC}$, whereas the conventional regularization fails and exhibits intermediate $\mathrm{uSC}$ and $\mathrm{2SC}_{us}$ windows. The 2SC to normal-conducting (NQ) boundaries of the minimal and massless schemes coincide. In the next section, we show how our results can be understood analytically in the large-density regime by approaching the symmetric massless CFL limit.
\section*{Analytic studies}
Here we present analytic results for the diquark gaps and speed of sound at $T=0$ and the critical temperatures $T_c$ in the symmetric massless CFL limit of the RG-consistent NJL model, where all diquark gaps are equal, \(\Delta_{ud}=\Delta_{us}=\Delta_{ds}\equiv\bar\Delta\). The nontrivial-solution (gap) condition at \(T=0\) in the minimal scheme reads
\begin{align}\label{eq:gapeqminimal}
0
&=
\frac{3}{2G_D}
 - \frac{2\mu^{2}}{\pi^{2}}\Big( -9 + \ln\!\frac{16\Lambda'^{6}}{\bar\Delta_{ }^{6}} \Big)\nonumber\\&\quad-\frac{2}{\pi^{2}}\Big( 2\Lambda'\sqrt{\bar\Delta_{ }^{2}+\Lambda'^{2}} + \Lambda'\sqrt{4\bar\Delta_{ }^{2}+\Lambda'^{2}} - 2\bar\Delta_{ }^{2}\big( 2\,\operatorname{arccsch}\!\frac{2\bar\Delta_{ }}{\Lambda'} + \operatorname{arsinh}\!\frac{\Lambda'}{\bar\Delta_{ }} \big) \Big).
\end{align}
This equation shows that the value of the diquark condensate saturates to a constant value at $\mu\to\infty$. For the minimal scheme this value is $\bar\Delta^\text{(min)}(\mu\to\infty)=\frac{2^{2/3}\,\Lambda'}{e^{3/2}}$, being $2^{-1/3}$ times the asymptotic value of the 2SC diquark condensate in Ref.~\cite{Gholami:2025afm}.
Similar expressions can be calculated for the massless scheme, giving a different asymptotic value for the gap, as seen in left panel of Fig.~\ref{fig:gaps}. The speed of sound squared in the minimal scheme, at very high $\mu$ is
\begin{align}\label{eq:cs2}
     c_s^2=(2\bar\Delta^{2}+\mu^{2})/(2\bar\Delta^{2}+3\mu^{2}).
\end{align}
A similar expression can be derived for the massless scheme. For both schemes, it shows that $c_s^2$ approaches the conformal limit from above for nonzero diquark gaps. The solution $\Delta(\mu)$ of Eq.~\eqref{eq:gapeqminimal}, Eq.~\eqref{eq:cs2} and their analog for the massless scheme are shown in Fig.~\ref{fig:gaps} left and right, respectively. For $\mu\gtrsim 600~\mathrm{MeV}$ the RG-consistent curves are indistinguishable from the symmetric-CFL calculation. For all schemes, the critical temperature of the 2SC-NQ transition in the massless symmetric limit, shown as a gray line in Fig.~\ref{fig:pd}, is
\begin{align}
T_c(\mu)\label{eq:tcNJL2}
= \frac{\sqrt{3}\,\mu}{\pi}
  \sqrt{\,
    W\!\left(
      \frac{4\,\Lambda'^{2}}{3\mu^{2}}
      \exp\!\left(
        -3 + 2\gamma_{\!E}
        + \frac{1}{\mu^{2}}
          \left(
            \Lambda'^{2}
            - \frac{\pi^{2}}{4G_D}
          \right)
      \right)
    \right)
  },
\end{align}
where \(W(x)\) denotes the Lambert \(W\) function and {$\gamma_E\approx0.577$ is the Euler-Masceroni constant. The ratio of $T_c$ to the diquark gap at $T=0$ approaches to
\(
  \lim_{\mu \to \infty} \frac{T_c(\mu)}{\bar\Delta_{}^{\text{(min)}}(\mu)} 
  \;=\; 2^{1/3}\,\frac{e^{\gamma_{\!E}}}{\pi}
  \;\approx\; 0.714
\)
for the minimal scheme in the limit $\mu\to\infty$. This is \(2^{1/3}\) times the BCS ratio \(e^{\gamma_{\!E}}/\pi \approx 0.567\). Again, we find very good agreement with the numerical result at high $\mu$, see Fig.~\ref{fig:pd}.

We conclude that physically reliable thermodynamics in the NJL model, especially for astrophysical applications, requires an RG-consistent treatment. Applications and implications of this framework are presented in Refs.~\cite{Gholami:2024ety,Christian:2025dhe}.

{\it Acknowledgment.}
This work was supported by the Deutsche Forschungsgemeinschaft (DFG, German Research Foundation) – project number 315477589 – TRR 211.
M.H. is supported by the GSI
F\&E.

\bibliography{bib.bib}

\end{document}